\documentclass[aps,prl,reprint,superscriptaddress]{revtex4-1}

\usepackage{graphicx}
\usepackage{dcolumn}
\usepackage{bm}

\begin{document}


\title{Plasmon mode modifies the elastic response of a nanoscale charge density wave system}



\author{Shamashis Sengupta}
\email[]{shamashis@tifr.res.in}
\affiliation{Department of Condensed Matter Physics and Materials Science, Tata Institute of Fundamental Research, Mumbai 400005, India
}
\author{Niveditha Samudrala}
\affiliation{Department of Condensed Matter Physics and Materials Science, Tata Institute of Fundamental Research, Mumbai 400005, India
}
\author{Vibhor Singh}
\affiliation{Department of Condensed Matter Physics and Materials Science, Tata Institute of Fundamental Research, Mumbai 400005, India
}
\author{Arumugam Thamizhavel}
\affiliation{Department of Condensed Matter Physics and Materials Science, Tata Institute of Fundamental Research, Mumbai 400005, India
}
\author{Peter B. Littlewood}
\affiliation{Physical Sciences and Engineering Division, Argonne National Laboratory, Argonne, Illinois 60439, USA
}
\author{Vikram Tripathi}
\email[]{vtripathi@theory.tifr.res.in}
\affiliation{Department of Theoretical Physics, Tata Institute of Fundamental Research, Mumbai 400005, India
}
\author{Mandar M. Deshmukh}
\affiliation{Department of Condensed Matter Physics and Materials Science, Tata Institute of Fundamental Research, Mumbai 400005, India
}


\date{\today}

\begin{abstract}
The elastic response of suspended NbSe$_3$ nanowires is studied across the charge density wave phase transition. The nanoscale dimensions of the
resonator lead to a large resonant frequency ($\sim$ 10-100 MHz), bringing the excited phonon frequency in close proximity of the
plasmon mode of the electronic condensate - a parameter window not accessible in bulk systems. The interaction between the phonon and plasmon modes strongly modifies the elastic properties at high frequencies. This
is manifested in the nanomechanics of the system as a sharp peak in the temperature dependence of the elastic modulus (relative change of 12.8$\%$) in the charge
density wave phase.
\end{abstract}


\maketitle

Research on nanoelectromechanical systems (NEMS) \cite{sazonova, poot} has progressed a lot over the last decade, providing
a way to explore the elastic properties of nanoscale systems. However, the ability of NEMS to act as a probe of the lattice
dynamics of a correlated system has not yet been widely recognized. This should be interesting because the screening of charge
by electrons strongly affects the lattice stiffness in a metal. \cite{ibach_luth}
The elastic properties are related to the dielectric constant - a frequency dependent quantity, and may be modified significantly
when the electrons undergo a phase transition to a coherent state. NEMS fashioned out of charge density wave (CDW) materials \cite{gruner1, gruner_zettl} provide the right
parameter window to combine the advantages of a high mechanical resonant frequency (in the MHz range) and an overlapping low
frequency collective plasmon mode. The CDW systems studied in our experiment are nanowires of the quasi-one-dimensional
material NbSe$_3$. Unlike bulk crystals, \cite{brillandong} the elastic modulus of these high frequency nanomechanical resonators shows an abrupt
peak with the variation of temperature in the CDW phase. Using a model of an incommensurate CDW interacting with the ionic
lattice, we attribute this surprising behaviour to a mutual locking of the lattice and the CDW. We discuss why this
effect is visible only at the nanoscale.

\begin{figure}
\includegraphics[width=80mm, bb=0 0 1037 936]{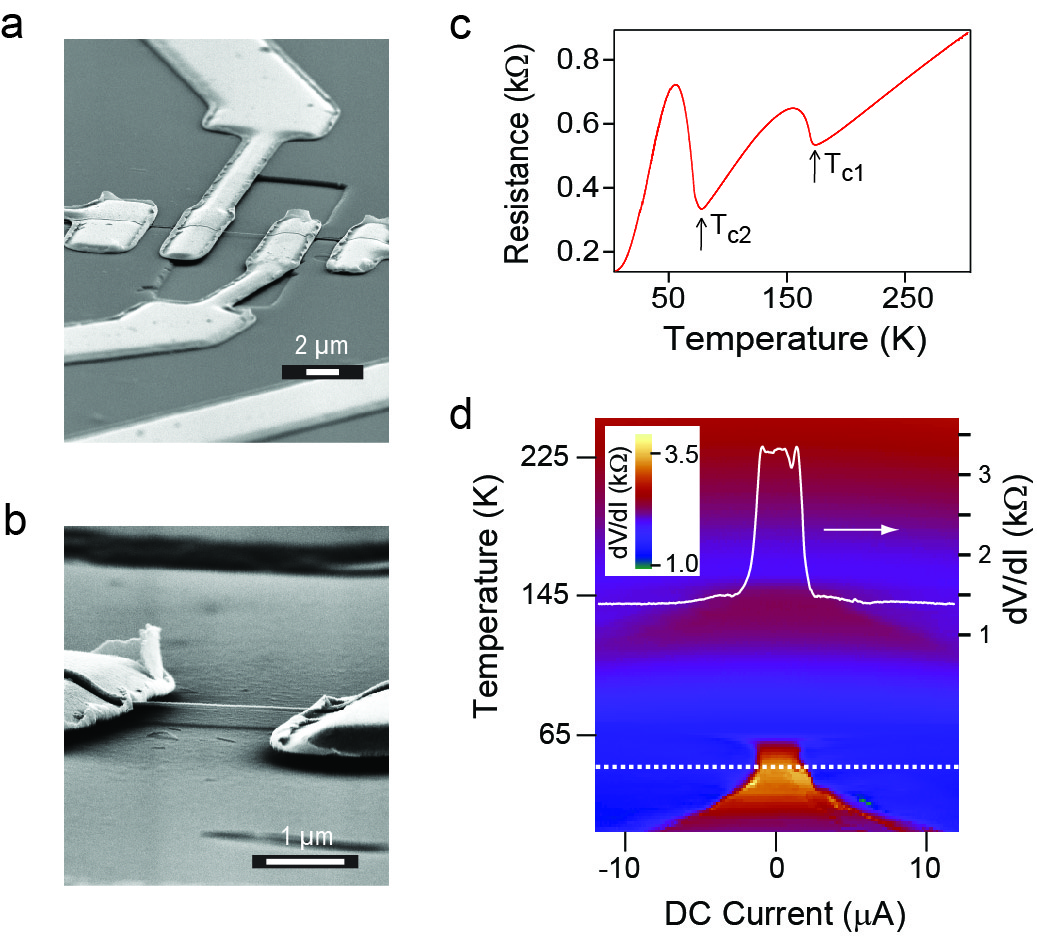}
\caption{\label{fig:figure1} (a), (b) Scanning electron microscope (SEM) images of suspended NbSe$_3$ nanowires. (c) Four probe measurement of resistance as a function of temperature for a nanowire. (d) The CDW depins and participates in conduction under the application of a large electric field. This is observed by ramping up the dc current through the device and monitoring the differential resistance ($dV/dI$, where $V$ is the voltage drop across the device and $I$ is the current flowing through it). Above a certain critical current, the electric field is sufficient to depin the CDW and the differential resistance drops sharply. A lineplot at 46.8 K is shown (indicated by dotted line). The sharp drop indicates the depinning of the CDW. The threshold electric field for depinning is estimated to be 11.4 V/cm. (Length of the nanowire = 3.1 $\mu$m.).}
\end{figure}

Crystals with a quasi-one-dimensional structure develop a charge density wave phase \cite{gruner1, gruner_zettl}
at low temperatures where the electron density assumes a periodic variation over space along with a distortion of the
lattice. The signatures of the new order parameter are observed in the electronic properties \cite{transport, sliding}
and concomitant changes in the elastic properties \cite{brillandong, brill_depin, sherwin} have been seen in macroscopic
crystals. In recent years, electrical transport measurements on NbSe$_3$ nanowires have shown that the properties of the
CDW phase are significantly different from macroscopic samples. \cite{slot1, crabtree, stabile} The behaviour of elastic
properties of nanoscale CDW systems are yet to be investigated in detail. Some measurements on thin flakes of the
layered CDW material NbSe$_2$ have been reported. \cite{nbse2} But the quasi-two-dimensional CDW phenomenon associated with this system is not well understood and a coherent picture of the underlying physics is missing. The CDW transition in the quasi-one-dimensional CDW system  NbSe$_3$ involves the gapping of a large part of the Fermi surface \cite{transport}, leading to a very prominent influence on its electrical transport properties. In this paper, we will probe the influence of the CDW on the elastic modulus and mechanics of nanowires of NbSe$_3$. The coherent charge density wave has a considerable effective mass \cite{littlewood87} and combined with the large dielectric constant of NbSe$_3$, \cite{gruner_RF} reduces the CDW plasmon frequency low enough to leave its signature on nanomechanical measurements.

Resonator devices were fabricated on a degenerately doped Si substrate (coated with a 300 nm dielectric layer of SiO$_2$). The nanowires were suspended in doubly clamped geometry with metallic contacts at both ends \cite{supp} (Figs. \ref{fig:figure1}(a) and \ref{fig:figure1}(b)). These were capacitively driven and set into mechanical vibrations to determine the resonant frequency. The change in elastic modulus was determined by measuring the resonant frequency of these devices across the charge density wave phase transition. Typically, the width of the contacted nanowires are between 80-300 nm and the thickness is between 30-100 nm. The length of the suspended part between the contacts varies between 2-4 $\mu$m. The result of four-probe measurement of the resistance ($R$) as a function of temperature ($T$) curve of one such nanowire is shown in Fig. \ref{fig:figure1}(c). The unit cell of NbSe$_3$ consists of three types of molecular chains, two of which undergo a CDW transition. There are two CDW transitions observed in NbSe$_3$ (marked T$_{c1}$ and T$_{c2}$ in Fig. \ref{fig:figure1}(c)). \cite{transport}

\begin{figure}
\begin{center}
\includegraphics[width=75mm, bb=0 0 1417 2671]{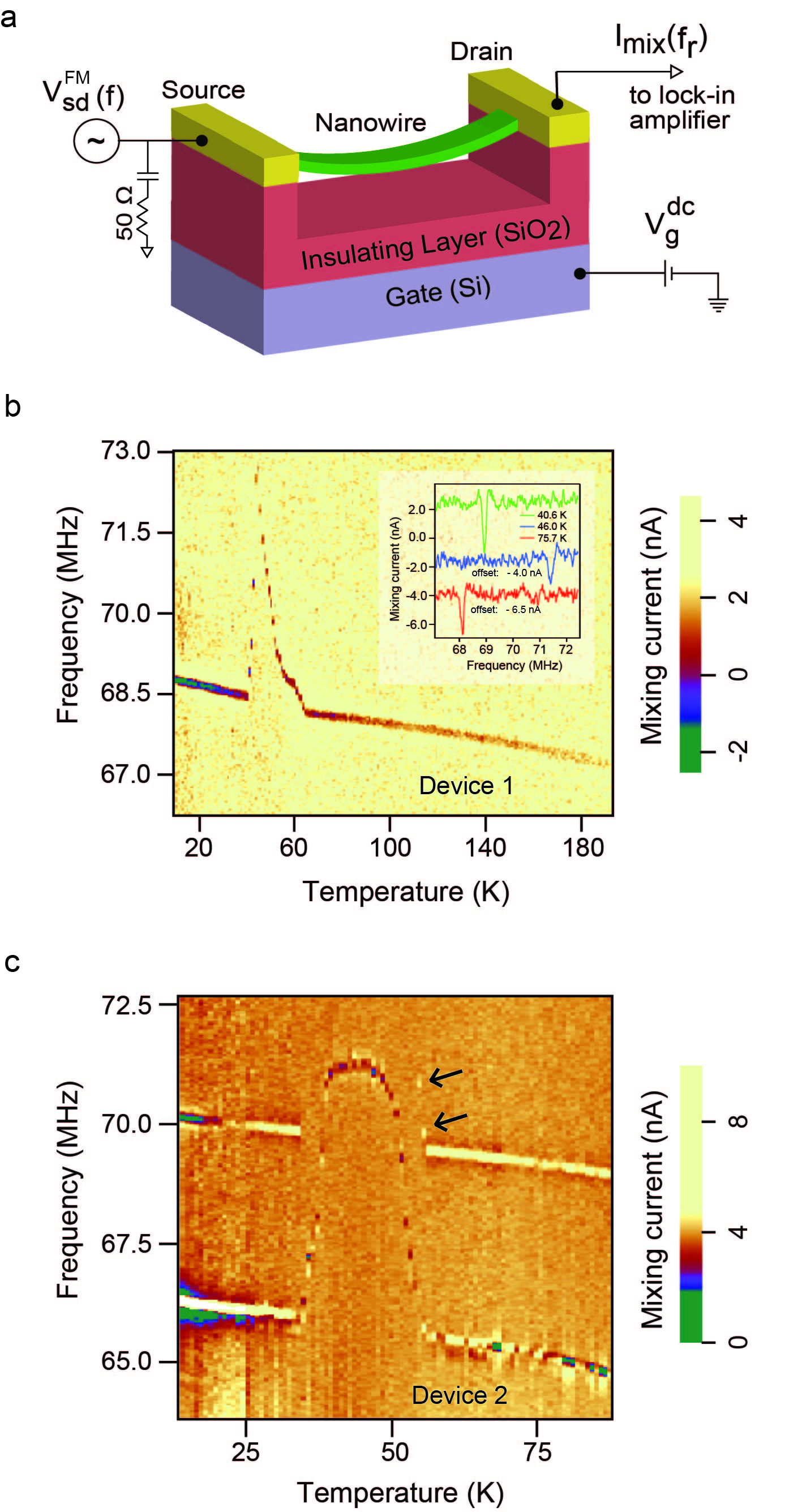}
\caption{\label{fig:figure2} (a) Schematic diagram of the circuit used for actuation and detection of mechanical resonance of a suspended nanowire. (b) Mixing current as a function of driving frequency and temperature for Device 1 ($V^{dc}_g$ = -44 V). There is a large peak in resonant frequency (and also in the elastic modulus) below 60 K - where a charge density wave transition takes place. (Inset: Lineplots of mixing current as a function of driving frequency at three different temperatures. The shift in the position of the peak (in `frequency' axis) for the curve at temperature of 46.0 K compared to the other two curves corresponds to the peak seen in the main colourscale plot. The curves are plotted with an offset (along the `mixing current' axis) for clarity.) (c) Variation of resonant frequency with temperature for Device 2 ($V^{dc}_g$ = -45 V).}
\end{center}
\end{figure}

Incommensurate CDWs (for which the CDW wavelength is not an integral multiple of the lattice spacing) \cite{fleming_xray} can slide in the presence of an electric field. The CDW is usually pinned by impurities but beyond a certain critical electric field (or a dc current passing through it), it depins and participates in electrical conduction. \cite{washboard, sliding} Differential resistance measurements (Fig. \ref{fig:figure1}(d)) on a nanowire device indicates the threshold electric field ($E_{th}$) for de-pinning of the CDW to be 11.4 V/cm at 46.8 K. This is 3 orders of magnitude larger than macroscopic crystals. \cite{sliding} A high depinning field as compared to bulk crystals has been reported to arise on the reduction of the size of the crystal. \cite{slot_prb, crabtree} The crossover lengthscale is determined by the phase coherence length \cite{slot_prb, thorne} of the CDW ($\sim$ few micrometers). At low temperatures (below 40 K), the resistance measurement plot in Fig. \ref{fig:figure1}(c) shows metallic behaviour (resistance decreasing upon the reduction of temperature). However, this is not the case for all our nanowires, and some of them show non-metallic \cite{slot1} behaviour at low temperatures (see Fig. S1 of Supplemental Material \cite{supp}). We now discuss the measurement of elastic properties by measuring the resonant frequency of suspended structures - the key aspect of our work.

NEMS devices constitute an active field of modern day research and have been used in diverse applications like ultra-sensitive mass detection, \cite{bachtold}, realizing bit-storage operations \cite{yamaguchi}, and studying the quantum ground state of a harmonic oscillator \cite{oscillatorstate}. The schematic diagram of our nanowire resonator device is shown in Fig. \ref{fig:figure2}(a). A frequency-modulated voltage $V^{FM}_{sd}$ at frequency $f$, deviation $f_\Delta$ and modulation rate $f_r$ is applied at the source and a dc voltage $V^{dc}_g$ is applied at the gate electrode. \cite{fm} (See Figs. S2 and S3 of Supplemental Material and the associated discussion about details of the measurement technique. \cite{supp}) $V^{FM}_{sd}(t)= V_0$cos$(2 \pi f t + \frac{f_\Delta}{f_r}$sin$(2 \pi f_r t))$, where $t$ denotes time and $V_0$ is the amplitude of the ac signal. This produces a driving force on the nanowire and sets it into oscillation at the same frequency $f$. The current through the device has a low frequency component at frequency $f_r$ ($\sim$ few hundred Hz) which shows a sharp change when the driving frequency $f$ of $V^{FM}_{sd}$ matches the natural resonant frequency of the nanowire. This component is called the mixing current $I_{mix}$ and is measured using a lock-in amplifier. $I_{mix} \propto \frac{dG}{dq} f \frac{(f^2-f_0^2+\frac{f_0^2}{Q})(f^2-f_0^2-\frac{f_0^2}{Q})}{((f_0^2-f^2)^2 + (\frac{f f_0}{Q})^2)^2}$, where $G$ is the conductance, $q$ the capacitively induced charge, $f_0$ the natural resonant frequency of the beam and $Q$ is the quality factor of resonance. At the resonant frequency, the amplitude of mechanical oscillation is maximum which leads to a sharp change of the mixing current. Therefore, by monitoring the mixing current as a function of frequency, we can determine the resonant frequency and from this the elastic modulus is inferred. The natural resonant frequency $f_0$ of a beam depends upon the elastic modulus $E$ and strain $\eta$ as $f_0 = \sqrt{\frac{EI}{\rho S}}(\frac{1}{2L}\sqrt{\eta \frac{S}{I}} + \frac{1}{L^2})$. ($L$ is the length of the suspended nanowire, $I$ the inertia moment, $\rho$ the mass density and $S$ is the cross-sectional area. \cite{sapmaz})

To measure the change of elastic modulus at the CDW phase transition, the resonant frequency is determined by sweeping the frequency of $V^{FM}_{sd}$ while measuring the mixing current, and the system is slowly heated up across the phase transition. The result from one device (Device 1) is shown in the colourscale plot of $I_{mix}$ in Fig. \ref{fig:figure2}(b). The positions of sharp change in $I_{mix}$ trace out the variation of resonant frequency. There is a prominent peak (between 40-63 K) observed just below the T$_{c2}$ CDW transition. Over the wide range of temperature from 10 K to 190 K, the resonant frequency has a general trend of decreasing monotonically with increasing temperature, except that there is the abrupt peak between 40-63K. The monotonically decreasing trend can be understood by considering the strain in the system. When the device is fabricated at room temperature, strain in the system should be small. On cooling it down, the nanowire as well as the metal electrodes holding it contract. This tends to pull the nanowire and increases the strain and leads to higher resonant frequencies at low temperature, the change being a gradual one. By a similar argument, when the system is heated up starting at 10 K, the nanowire and the electrodes will expand, thus reducing the strain and also the resonant frequency. This explains the gradual decrease of resonant frequency with increase in temperature. Now, in the region of the peak (at 40 K), the resonant frequency goes up abruptly. This can not be due to strain resulting from thermal expansion, because as explained earlier, strain would reduce with increasing temperature and will not lead to an increase in $f_0$. The quantity, a change in which can give rise to the `peak', is the elastic modulus. Considering that $f_0 \propto \sqrt{E}$, the height of the `peak' corresponds to a fractional change in elastic modulus of 12.8$\%$ ($\pm$0.9$\%$). This is an extremely large change. Comparison of the data of Fig. \ref{fig:figure2}(b) with elastic modulus data on macroscopic crystals yields a striking contrast. In the vibrating reed experiments on bulk crystals by Brill et al., \cite{brillandong} there was no detectable change in elastic modulus at the T$_{c2}$ CDW transition around 60 K. However, a `dip' corresponding to a 0.09$\%$ change in elastic modulus was seen near the T$_{c1}$ transition at 142 K. We do not detect any significant feature around the T$_{c1}$ transition. (Our devices have Q$\sim$300 and this may not be sufficient for detecting the very small change in elastic modulus reported by Brill et al. \cite{brillandong} Further discussions are provided in Supplemental Material. \cite{supp})

Fig. \ref{fig:figure2}(c) shows data from a second device (Device 2). Two resonant modes can be noticed in this plot. Closely spaced modes arise in NEMS devices because of small asymmetries in device geometry. \cite{hari} For the upper mode, the signal of resonance becomes extremely faint as the resonant frequency starts increasing sharply beyond 35 K and the `peak' feature can not be traced (except for two points on the right hand `slope' of the peak indicated by two arrows in the figure). However, the feature is clear for the lower mode and it is estimated that the maximum value of the `peak' in resonant frequency corresponds to a 16.2 $\%$ change in the elastic modulus. (Data from one other device is presented in Fig. S4 of Supplemental Material. \cite{supp} Measurements carried out on a total of six devices showed similar qualitative features.)

In the following part of the paper, we discuss the mechanism that leads to the abrupt variation of resonant frequency in our nanowire devices near the T$_{c2}$ transition, and also why such a feature was absent in measurements on bulk crystals. Phonons in metals arise from a Bohm-Staver mechanism where the restoring force for lattice displacements is provided by screening effects of the electron gas. In a simple picture,
an one-dimensional lattice can be thought of as a chain of masses connected by springs, each of stiffness $K_s$ and the stiffness scales inversely with the dielectric constant $\epsilon$ ($K_s\propto\frac{1}{\epsilon}$). \cite{ibach_luth} In the case of the CDW system NbSe$_3$, the lattice also sees the potential due to the incommensurate charge density wave. The potential energy of the lattice,

\begin{equation}
\Phi=\sum_i \frac{1}{2}K_s(x_i-x_{i-1}-l)^2 + \Gamma (1-\cos(2 \pi \frac{x_i}{\lambda}))\;.
\label{eq:equation3}
\end{equation}

\begin{figure}
\begin{center}
\includegraphics[width=80mm, bb=0 0 873 821]{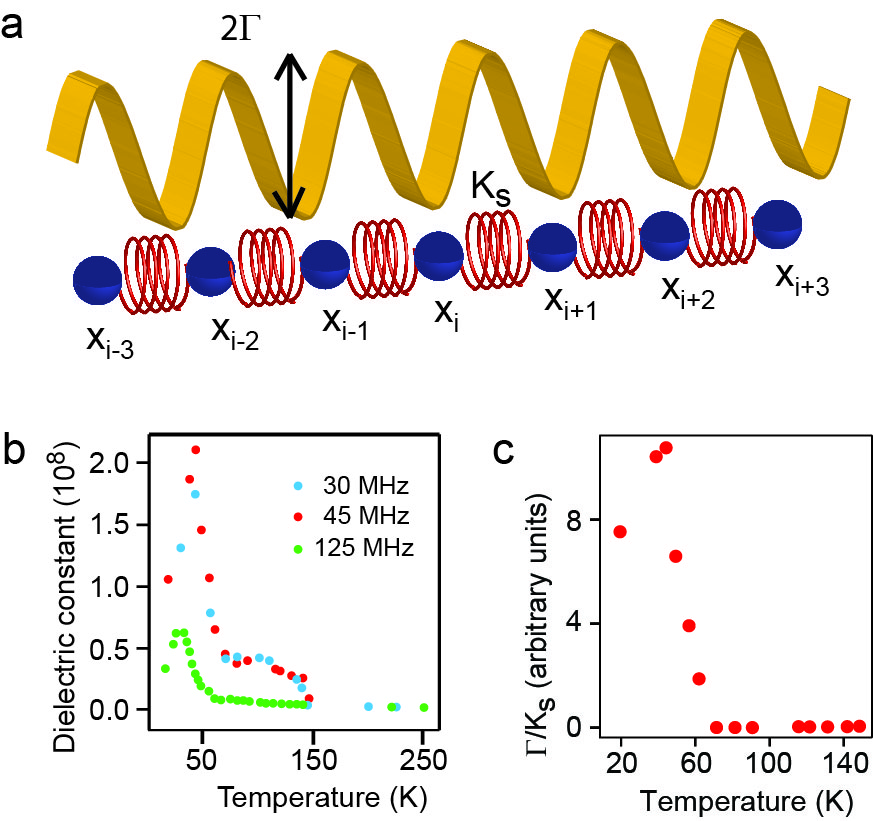}
\caption{\label{fig:figure3} (a) The lattice is represented by a chain of masses connected by springs each with stiffness $K_s$. The position of the $i^{th}$ ion is given by $x_i$. The potential of the CDW is periodic with an amplitude $\Gamma$. $K_s \sim \frac{1}{\epsilon}$, where $\epsilon$ is the dielectric constant of the system. (b) Data of dielectric constant of NbSe$_3$ at RF frequencies as a function of temperature reported by Gruner et al. \cite{gruner_RF} (c) The parameter $\Gamma/K_s$ is plotted as a function of temperature ($T$). It is evaluated following the relation $K_s\sim\frac{1}{\epsilon}$, with the data in (b) for 45 MHz frequency being used as a prototype for the variation of $\epsilon$ with $T$. The potential due to the CDW is assumed to follow a Landau-Ginzburg expression: $\Gamma \sim (T_{c2}-T)^{1/2}$.}
\end{center}
\end{figure}

Here, $\Gamma$ is proportional to the amplitude (or the order parameter) of the CDW, $x_i$ is the position of the $i^{th}$
ion, $l$ is the equilibrium separation between the ions and $\lambda$ is the wavelength of the CDW. (See Fig. \ref{fig:figure3}
(a) for a sketch of the model.) This is known in literature as the Frenkel-Kontorova model. \cite{aubry} In order to evaluate
the resonant frequency, one needs to estimate the phonon modes. It had been shown by numerical computations \cite{aubry} that when the ratio $\frac{\Gamma}{K_s}$ exceeds a critical value, a `phonon gap' opens up, shifting the acoustic phonon frequencies to larger values. Physically, the absence of low frequency modes at long wavelengths originates in the pinning of the lattice ions to the CDW potential. When the CDW potential is weak, the lattice vibrations are able to slide with respect to the CDW and one obtains the usual Bohm-Staver phonons. It had been observed in measurements of the dielectric constant of NbSe$_3$ at microwave frequencies that $\epsilon$ has a large peak around 42 K (Fig. \ref{fig:figure3}(b), data reproduced from Gruner et al. \cite{gruner_RF}). This implies that the parameter $\frac{\Gamma}{K_s}$ (Fig. \ref{fig:figure3}(c)) will have a peak in the similar temperature range - presumably taking the system into the `phonon gap' regime producing a large change in the experimentally measured resonant frequency. In Eq. (\ref{eq:equation3}), we have disregarded the vibrations of the CDW, assuming it to be of infinite mass. In reality, the CDW is coherent up to a length scale of a few microns and thus has a mass $m^{*}\sim 10^4 m_{e}$. \cite{littlewood87} As this is of the order of the ionic masses
that are associated with the phonons, we should also take into account the vibrations of the CDW as they will mix with the phonon modes.

There are two ways in which the CDW can contribute to the apparent stiffness of the sound modes. The most natural one is that the opening up of the CDW gap increases the stiffness of the electron gas, which itself contributes to the elastic modulus assuming that the lattice and CDW are in equilibrium. In general, this is a small effect, of the order of
$(\Delta/W)^2 \simeq 10^{-6}$, the squared ratio of the CDW gap ($\Delta$) to the overall electronic bandwidth ($W$). This is consistent with measurements on bulk samples. A second route is more subtle, and is associated with the fact that
the CDW, when distorted by being pinned to the lattice, produces long range Coulomb forces, which are themselves screened by the metallic background in equilibrium. However, when driven at a frequency which is too rapid for the electron gas to respond, the frequency of a longitudinal sound mode will be shifted up to the `CDW plasmon'.

The sinusoidal charge distribution over space $n(\textbf{r})$ is represented as $n(\textbf{r})=n_c+n_0 \cos [\textbf{Q.r} + \phi(\textbf{r})]$. $n_0$ is the amplitude and $\textbf{Q}$ is the wave-vector of the CDW. $n_c$ denotes the electron density in the metallic state. The spatial variation in the phase $\phi(\textbf{r})$ of the CDW results from the deformation of the CDW due to pinning and applied electric fields. The dynamics of the charge density wave can be understood by analyzing the equation of motion of the CDW phase $\phi(\textbf{r})$. \cite{littlewood87} The response of the system at frequency $f$ is given by

\begin{equation}
G_{L}^{-1} = -4\pi^2 m^{*}f^2 -i2\pi\gamma_0 f + V_0 - i2\pi f \frac{n_c e^2}{\sigma_z - i2\pi f \epsilon_0 \epsilon_z}\;,
\label{eq:equation7}
\end{equation}

where, $V_0$ is the CDW pinning potential coming from impurities and the lattice, $\gamma_0$ is the damping coefficient, $\sigma_z$ and $\epsilon_z$ denote the conductivity and dielectric constant respectively along the chain direction (along which the CDW can slide). We have assumed for simplicity that this pinning is uniform (see Ref. \onlinecite{littlewood87} for a discussion).  The last term of the above equation indicates that there exist two different frequency regimes where the dynamics of the CDW is distinctly different and a cross-over between these two regimes occur at a characteristic frequency of $f_c$=$\sigma_z/(2\pi\epsilon_0 \epsilon_z)$. At low frequencies ($f\ll\frac{\sigma_z}{2\pi\epsilon_0 \epsilon_z}$), the response is overdamped, with an effective damping $\gamma = \gamma_0 + n_c e^2/\sigma_z.$ In this regime, the motion of the CDW is strongly screened by the conduction electrons. Any effect of the coupling of the CDW with the lattice (such as in the elastic properties) is hard to detect given the broadness of the response. At high frequencies, $f\gg\frac{\sigma_z}{2\pi\epsilon_0 \epsilon_z}$,
the last term in Eq. (\ref{eq:equation7}) acts as a restoring term instead of a damping term, and one obtains a (CDW plasmon) resonance at a frequency $f_p$ given by

\begin{equation}
 f_{p} = \frac{1}{2\pi}\sqrt{V_0/m^{*} + n_c e^2/m^{*}\epsilon_0\epsilon_z}.
\label{eq:equation8}
\end{equation}

Using the conductivity data from electrical transport measurements on our devices and the dielectric constant values as measured by Gruner et al., \cite{gruner_RF} the crossover frequency $f_c = \frac{\sigma_z}{2 \pi \epsilon_0 \epsilon_z}$ is estimated to be $f_c\sim$20 MHz. \cite{supp} This shows that our devices with resonant frequencies of tens of MHz exceed this crossover. In contrast, experiments for determining the elastic modulus of macroscopic crystals by Brill et al. were performed with resonant frequencies not exceeding a few kHz (0.1-1.7 kHz). \cite{brillandong} It is not expected that the coupling of the CDW condensate to the mechanical motion of the ionic lattice would be discernible at such low frequencies.

Now we estimate the CDW plasmon frequency $f_p$. Assuming that the pinning potential can be neglected, $f_p \sim (1/2\pi)\sqrt{n_c e^2/m^{*}\epsilon_0 \epsilon_z}$. We take $n_c\sim$2$\times$10$^{25}$ m$^{-3}$ from an
analysis presented in Ref. \onlinecite{ong_twoband}. The effective mass of the CDW m$^*\sim$10$^4$m$_e$
(Ref. \onlinecite{littlewood87}). Using these numbers, we get $f_p\sim$39 MHz. \cite{supp} This is of the same order as the resonant frequencies of our nanowire devices, which means the Bohm-Staver phonons are close to the CDW plasmon frequency and the phonon modes can get  renormalized due to a coupling between the ionic lattice and the CDW. It has to be kept in mind that the estimate $f_p\sim$39 MHz is an approximate one. The value of $n_c$ used for the calculation is an order-of-magnitude estimate. \cite{ong_twoband} We have taken $m_e$ to be the bare electron mass. It should be the effective mass of electrons in NbSe$_3$, and was determined to be of the order of unity from magnetotransport measurements \cite{monceau_sdh}. The exact value of $f_p$ may be larger (though the order of magnitude should be the same) than the estimate provided. Thus, upon the renormalization of the phonon modes due to the closeness to the CDW plasmon mode, the elastic modulus of the NbSe$_3$ nanowires can undergo a sharp increase as has been experimentally seen in our devices.


On the basis of the Frenkel-Kontorova \cite{aubry} model and the measurements of the high frequency dielectric constant by Gruner
et al. \cite{gruner_RF}, the following picture emerges for our experiments. For temperatures $T>>42\rm{K},$ the resonance frequency corresponds to the longitudinal phonons corresponding
to a wavelength of the order of the NbSe$_3$ sample. In the Frenkel-Kontorova model, \cite{aubry} this regime corresponds to
free lattice vibrations unrestrained by the CDW (which is either weak or nonexistent). At lower temperatures, when the CDW
transition takes place, the dielectric function is experimentally known to show a strong enhancement when measured at
microwave frequencies of the order of 100MHz \cite{gruner_RF}. This softens the Bohm-Staver phonons which now get locked to
the CDW potential resulting in the opening of a phonon gap at long wavelengths. As the temperature is lowered further, the
dielectric function again decreases \cite{gruner_RF} and the lattice vibrations are no longer locked to the CDW.

This Letter demonstrates that the elastic properties of nanoscale resonators fabricated using a charge density wave material are
strongly affected by the plasmon mode of the electronic condensate. This provides us with a new window for interesting
explorations of the coupling between electronic order and the mechanics of nanostructures. There exist
a number of correlated solid state systems with intriguing phase transition behaviours involving interactions between the
electrons and the lattice - such as materials with a structural Mott transition \cite{wu} and vortex phases in the peak
effect regime of superconducting NbSe$_2$. \cite{peakeffect} Probing at the nanoscale can also highlight the behaviour of
individual phase domains. This is a less explored avenue and leaves a great scope for future work on NEMS.

We thank S. Bhattacharya, P. Monceau, H. S. J. van der Zant and U. Waghmare for discussions.
Work done at TIFR is supported by the Government of India. Work done at Argonne is supported by the US Department of Energy under FWP 70069.


\end{document}